\documentclass[twocolumn,superscriptaddress,nofootinbib]{revtex4-1}
\usepackage{amsmath,amssymb,bm,graphicx,hyperref}
\allowdisplaybreaks[1]

\renewcommand\d{\partial}
\newcommand\+{\dagger}

\newcommand\0{{\bm{0}}}
\newcommand\p{{\bm{p}}}
\newcommand\x{{\bm{x}}}
\newcommand\y{{\bm{y}}}
\newcommand\C{\mathcal{C}}
\newcommand\E{\mathcal{E}}
\renewcommand\H{\mathcal{H}}
\newcommand\M{\mathcal{M}}
\renewcommand\O{\mathcal{O}}
\renewcommand\P{\mathcal{P}}
\newcommand\Q{\mathcal{Q}}
\renewcommand\S{\mathcal{S}}
\newcommand\T{\mathcal{T}}
\newcommand\eq{\mathrm{eq}}
\newcommand\n{\mathrm{n}}
\newcommand\s{\mathrm{s}}
\newcommand\Tr{\mathrm{Tr}}

\begin{document}

\title{Hydrodynamics with spacetime-dependent scattering length}

\author{Keisuke Fujii}
\affiliation{Department of Physics, Tokyo Institute of Technology,
Ookayama, Meguro, Tokyo 152-8551, Japan}
\affiliation{Interdisciplinary Theoretical and Mathematical Sciences Program, RIKEN,
Hirosawa, Wako, Saitama 351-0198, Japan}
\author{Yusuke Nishida}
\affiliation{Department of Physics, Tokyo Institute of Technology,
Ookayama, Meguro, Tokyo 152-8551, Japan}

\date{July 2018}

\begin{abstract}
Hydrodynamics provides a concise but powerful description of long-time and long-distance physics of correlated systems out of thermodynamic equilibrium.
Here we construct hydrodynamic equations for nonrelativistic particles with a spacetime-dependent scattering length and show that it enters constitutive relations uniquely so as to represent the fluid expansion and contraction in both normal and superfluid phases.
As a consequence, we find that a leading dissipative correction to the contact density due to the spacetime-dependent scattering length is proportional to the bulk viscosity ($\zeta_2$ in the superfluid phase).
Also, when the scattering length is slowly varied over time in a uniform system, the entropy density is found to be produced even without fluid flows in proportion to the bulk viscosity, which may be useful as a novel probe to measure the bulk viscosity in ultracold-atom experiments.
\end{abstract}

\maketitle
\tableofcontents

\section{Introduction}
Ultracold atoms provide versatile platforms to investigate various aspects of correlated systems both in and out of thermodynamic equilibrium~\cite{Dalfovo:1999,Giorgini:2008,Bloch:2008,Polkovnikov:2011,Langen:2015}.
Here the quantum statistics of particles is controlled by the choice of atomic isotopes and dimensionality of space by the application of optical lattices~\cite{Bloch:2008}.
In addition, an interparticle interaction is not only tunable in its magnitude and sign with the magnetic field via Feshbach resonances~\cite{Chin:2010}, but also variable over space and time at will to a reasonable extent~\cite{Pollack:2010,Clark:2017,Arunkumar:arxiv}.
While such a spacetime-dependent scattering length has been proposed to realize a number of intriguing phenomena~\cite{Rodas-Verde:2005,Balbinot:2008,Salerno:2008,Bulgac:2009,Gong:2009,Nishida:2012,Rapp:2012,Liberto:2014,Greschner:2014,Langmack:2015,Smith:2015}, it may also be useful as a novel probe of target systems.

The purpose of this paper is to shed light on possible roles of the spacetime-dependent scattering length by employing a hydrodynamic description of correlated systems with contact interactions.
To this end, we first derive a set of operator identities involving conserved charge and current densities in Sec.~\ref{sec:microscopics} by allowing the scattering length to be spacetime dependent.
Hydrodynamic constitutive relations are then constructed for normal fluids in Sec.~\ref{sec:normal-fluids} and for superfluids in Sec.~\ref{sec:superfluids} by imposing the second law of thermodynamics.
Here the spacetime-dependent scattering length proves to enter uniquely so as to represent the fluid expansion and contraction and thus be coupled with the bulk viscosities.
We finally conclude this paper in Sec.~\ref{sec:conclusion} with possible implications of our findings for ultracold-atom physics.
Some of our outcomes are also confirmed microscopically in Appendix~\ref{sec:appendix} without relying on the hydrodynamics.

In what follows, we will set $\hbar=k_B=1$ and employ shorthand notations $(x)=(t,\x)$ for spacetime coordinates and $\phi\tensor\d_{\!\mu}\psi\equiv[\phi(\d_\mu\psi)-(\d_\mu\phi)\psi]/2$ with $\mu=t$ or $i$.
Space indices are represented by $i=1,2,\dots,d$ and spin indices by $\sigma=1,2,\dots,N$.
Unless otherwise specified, implicit sums over repeated indices are assumed as well as for $[v_i]^2\equiv v_iv_i$.

\section{Quantum field theory}\label{sec:microscopics}
\subsection{Hamiltonian and equation of motion}
Let us consider nonrelativistic bosons or fermions with $N$ spin components in an arbitrary spatial dimension $d$, whose Hamiltonian is provided by
\begin{align}
\hat{H}(t) &= \int\!d^d\x\biggl[\frac{D_i\hat\psi_\sigma^\+(x)D_i\hat\psi_\sigma(x)}{2m}
- A_t(x)\hat\psi_\sigma^\+(x)\hat\psi_\sigma(x) \notag\\
&\quad + \frac{\lambda(x)}{2m}\hat\psi_\sigma^\+(x)\hat\psi_\tau^\+(x)
\hat\psi_\tau(x)\hat\psi_\sigma(x)\biggr].
\end{align}
Here $D_\mu\equiv\d_\mu-iA_\mu(x)$ is the covariant derivative and an external gauge field $A_\mu(x)$ is introduced for generality, whose temporal component is nothing short of a trapping potential and spatial components are produced in noninertial frames of reference.
In addition, $\lambda(x)$ is a spacetime-dependent bare coupling for the U($N$)-symmetric contact interaction.
In the dimensional regularization, it is related to the scattering length $a(x)$ via
\begin{align}\label{eq:scattering}
\lambda(x) = (d-2)\Omega_{d-1}a^{d-2}(x),
\end{align}
where $\Omega_{d-1}\equiv(4\pi)^{d/2}/2\Gamma(2-d/2)=2,2\pi,4\pi$ coincides with the surface area of the unit $(d-1)$-sphere for $d=1,2,3$.%
\footnote{The two-body scattering $\T$ matrix in the center-of-mass frame is provided by
$\T^{-1}(E)=\frac{m}{\lambda}-\int\!\frac{d^d\p}{(2\pi)^d}\frac{m}{mE-\p^2+i0^+}
=\frac{m}{\lambda}-\frac{m\kappa^{d-2}}{(d-2)\Omega_{d-1}}\big|_{\kappa\equiv\sqrt{-mE-i0^+}}$,
where the integral is analytically continued to an arbitrary $d$ after evaluated for $0<d<2$.
Therefore, the scattering length in Eq.~(\ref{eq:scattering}) is defined so that the two-body bound state existing for a constant $a>0$ has its binding energy at $E=-1/ma^2$.
See also Ref.~\cite{Valiente:2012} for a different convention.}

The annihilation operator $\hat\psi_\sigma(x)$ satisfies the equal-time commutation or anticommutation relation
\begin{align}
[\hat\psi_\sigma(t,\x),\hat\psi_\tau^\+(t,\y)]_\pm = \delta_{\sigma\tau}\delta^d(\x-\y)
\end{align}
and its time evolution is governed by the Heisenberg equation of motion
\begin{align}\label{eq:heisenberg}
& i\d_t\hat\psi_\sigma(x) = [\hat\psi_\sigma(x),\hat{H}(t)] \notag\\
&= \biggl[-\frac{\Delta}{2m} - A_t(x)
+ \frac{\lambda(x)}{m}\hat\psi_\tau^\+(x)\hat\psi_\tau(x)\biggr]\hat\psi_\sigma(x).
\end{align}
Here $\Delta\equiv D_iD_i$ is the Laplacian and the resulting Heisenberg equation is formally invariant under the local gauge transformation of
\begin{subequations}\label{eq:gauge}
\begin{align}
\hat\psi_\sigma(x) &\to e^{i\chi(x)}\hat\psi_\sigma(x), \\
A_\mu(x) &\to A_\mu(x) + \d_\mu\chi(x).
\end{align}
\end{subequations}

\subsection{Continuity equations}
Because of the gauge and spacetime-translational symmetries, the mass, momentum, and energy are intrinsically conserved up to external contributions of $A_\mu(x)$ and $a(x)$ with spacetime dependences.
The corresponding continuity equations follow straightforwardly from the Heisenberg equation (\ref{eq:heisenberg}), which involve the mass and momentum densities for each spin component (no implicit sums over $\sigma$)
\begin{align}\label{eq:operator_mass}
\hat\M_\sigma(x) &\equiv m\hat\psi_\sigma^\+(x)\hat\psi_\sigma(x), \\
\hat\P_{\sigma i}(x) &\equiv -i\hat\psi_\sigma^\+(x)\tensor{D}_i\hat\psi_\sigma(x),
\end{align}
as well as their totals $\hat\M(x)\equiv\sum_\sigma\hat\M_\sigma(x)$ and $\hat\P_i(x)\equiv\sum_\sigma\hat\P_{\sigma i}(x)$, the energy density without the trapping potential term
\begin{align}
\hat\H(x) \equiv \frac{D_i\hat\psi_\sigma^\+(x)D_i\hat\psi_\sigma(x)}{2m}
+ \frac{\lambda(x)}{2m}\hat\psi_\sigma^\+(x)\hat\psi_\tau^\+(x)\hat\psi_\tau(x)\hat\psi_\sigma(x),
\end{align}
the stress tensor
\begin{align}
& \hat\Pi_{ij}(x) \equiv \frac{D_i\hat\psi_\sigma^\+(x)D_j\hat\psi_\sigma(x)
+ D_j\hat\psi_\sigma^\+(x)D_i\hat\psi_\sigma(x)}{2m} \notag\\
& + \delta_{ij}\biggl[\frac{\lambda(x)}{2m}
\hat\psi_\sigma^\+(x)\hat\psi_\tau^\+(x)\hat\psi_\tau(x)\hat\psi_\sigma(x)
- \frac{\Delta[\hat\psi_\sigma^\+(x)\hat\psi_\sigma(x)]}{4m}\biggr],
\end{align}
the energy flux
\begin{align}\label{eq:operator_energy-flux}
\hat\Q_i(x) &\equiv \frac{D_i\hat\psi_\sigma^\+(x)\Delta\hat\psi_\sigma(x)
- \Delta\hat\psi_\sigma^\+(x)D_i\hat\psi_\sigma(x)}{4im^2} \notag\\
&\quad + \frac{\lambda(x)}{im^2}\hat\psi_\sigma^\+(x)
[\hat\psi_\tau^\+(x)\tensor{D}_i\hat\psi_\tau(x)]\hat\psi_\sigma(x),
\end{align}
and the so-called contact density~\cite{Braaten:2008}
\begin{align}\label{eq:operator_contact}
\hat\C(x) \equiv \frac{\lambda^2(x)}{2}\hat\psi_\sigma^\+(x)\hat\psi_\tau^\+(x)
\hat\psi_\tau(x)\hat\psi_\sigma(x).
\end{align}
In terms of the above local operators, the mass continuity equation is provided by
\begin{align}\label{eq:continuity_mass}
\d_t\hat\M_\sigma(x) + \d_i\hat\P_{\sigma i}(x) = 0,
\end{align}
the momentum continuity equation by
\begin{align}\label{eq:continuity_momentum}
& \d_t\hat\P_i(x) + \d_j\hat\Pi_{ij}(x) \notag\\
&= F_{it}(x)\frac{\hat\M(x)}{m} + F_{ij}(x)\frac{\hat\P_j(x)}{m}
- \frac{\d_ia(x)}{\Omega_{d-1}a^{d-1}(x)}\frac{\hat\C(x)}{m},
\end{align}
and the energy continuity equation by
\begin{align}\label{eq:continuity_energy}
\d_t\hat\H(x) + \d_i\hat\Q_i(x) = F_{it}(x)\frac{\hat\P_i(x)}{m}
+ \frac{\d_ta(x)}{\Omega_{d-1}a^{d-1}(x)}\frac{\hat\C(x)}{m},
\end{align}
where $F_{\mu\nu}(x)\equiv\d_\mu A_\nu(x)-\d_\nu A_\mu(x)$ is the field strength tensor.
The continuity equations are all gauge invariant and the right-hand sides of the momentum and energy continuity equations represent the external forces and powers supplied by $A_\mu(x)$ and $a(x)$.

In addition, the trace of the stress tensor proves to satisfy
\begin{align}\label{eq:operator_stress-trace}
\hat\Pi_{ii}(x) = 2\hat\H(x) + \frac{\hat\C(x)}{m\Omega_{d-1}a^{d-2}(x)}
- d\frac{\Delta\hat\M(x)}{4m^2},
\end{align}
which readily follows from the definitions in Eqs.~(\ref{eq:operator_mass})--(\ref{eq:operator_contact}) with Eq.~(\ref{eq:scattering}).
The resulting operator identity is the nonrelativistic counterpart of the tracelessness condition for conformality and the second term on the right-hand side thus provides the measure of conformal symmetry breaking~\cite{Hofmann:2012,Taylor:2012,Dusling:2013}.
We also note that the last term of Eq.~(\ref{eq:operator_stress-trace}) is not unique and can even be eliminated by redefining the stress tensor as $\hat\Pi_{ij}(x)\to\hat\Pi_{ij}(x)-\frac{d}{d-1}(\delta_{ij}\Delta-\d_i\d_j)\frac{\hat\M(x)}{4m^2}$, with the momentum continuity equation (\ref{eq:continuity_momentum}) kept intact.
Such an ambiguity at $O(\d^2)$ is however irrelevant to our discussion below as long as hydrodynamics up to first order in derivatives is concerned [see Eqs.~(\ref{eq:normal_stress-trace}) and (\ref{eq:super_stress-trace})].

\section{Hydrodynamics for normal fluids}\label{sec:normal-fluids}
\subsection{Constitutive relations}\label{sec:normal_constitutive}
When the system is perturbed out of thermodynamic equilibrium, its long-time and long-distance physics is governed by hydrodynamics founded on mass, momentum, and energy conservation laws as well as on local thermodynamic equilibrium.%
\footnote{In order for the thermodynamic limit to exist in the system of bosons, their interaction must be repulsive, which is possible for contact interactions only in one spatial dimension.}
The corresponding continuity equations follow from our operator identities in Eqs.~(\ref{eq:continuity_mass})--(\ref{eq:continuity_energy}) just by replacing each local operator therein with its expectation value denoted by $\O(x)\equiv\Tr[\hat\O(x)\hat\rho]$.
Here the density matrix $\hat\rho$ is arbitrary but independent of time because we work in the Heisenberg picture.

Hydrodynamics furthermore expresses the expectation values of the conserved charge and current densities in Eqs.~(\ref{eq:operator_mass})--(\ref{eq:operator_energy-flux}) in terms of the local thermodynamic variables and the fluid flow velocity $v_i(x)$.
The constitutive relations for normal fluids read
\begin{align}
\P_{\sigma i}(x) = \M_\sigma(x)v_i(x)
\end{align}
for the momentum densities,
\begin{align}\label{eq:normal_energy}
\H(x) = \E(x) + \frac{\M(x)}{2}[v_i(x)]^2
\end{align}
for the energy density,
\begin{align}\label{eq:normal_stress}
\Pi_{ij}(x) = p(x)\delta_{ij} + \M(x)v_i(x)v_j(x) + \pi_{ij}(x),
\end{align}
for the stress tensor, and
\begin{align}
\Q_i(x) = [\H(x)+p(x)]v_i(x) + q_i(x)
\end{align}
for the energy flux~\cite{Landau-Lifshitz}.
Here $\E(x)$ is the internal energy density and $p(x)$ is the pressure, while $\pi_{ij}(x)\,[{=}\,\pi_{ji}(x)]$ and $q_i(x)\sim O(\d)$ are the dissipative corrections to the stress tensor and the energy flux, respectively.

In addition, by substituting the constitutive relations for the energy density and the stress tensor in Eqs.~(\ref{eq:normal_energy}) and (\ref{eq:normal_stress}) into the expectation value of the operator identity in Eq.~(\ref{eq:operator_stress-trace}), we obtain
\begin{align}\label{eq:normal_stress-trace}
\frac{\C(x)}{m\Omega_{d-1}a^{d-2}(x)} = d{\,\cdot\,}p(x) - 2\E(x) + \pi_{ii}(x) + O(\d^2)
\end{align}
up to first order in derivatives.%
\footnote{Here and below, a dot emphasizing a product ($d{\,\cdot\,}\O$) is inserted after the spatial dimension $d$ to avoid confusion with a differential ($d\O$).}
Therefore, the contact density in local thermodynamic equilibrium is to be identified as
\begin{align}\label{eq:normal_pressure}
\frac{\C_\eq(x)}{m\Omega_{d-1}a^{d-2}(x)} \equiv d{\,\cdot\,}p(x) - 2\E(x),
\end{align}
which is the local extension of the thermodynamic identity known as the pressure relation~\cite{Tan:2008a,Tan:2008b,Tan:2008c,Braaten:2012}.
Here the equilibrium contact density $\C_\eq(x)$ is locally specified by $\M_\sigma(x)$, $\E(x)$, and $a(x)$ via the thermodynamic equation of state and should be distinguished from the genuine contact density $\C(x)\equiv\Tr[\hat\C(x)\hat\rho]$ not necessarily in local thermodynamic equilibrium.
Its constitutive relation is thus found to be
\begin{align}\label{eq:normal_contact}
\C(x) = \C_\eq(x) + m\Omega_{d-1}a^{d-2}(x)\pi_{ii}(x) + O(\d^2),
\end{align}
where the dissipative correction coincides with that to the stress tensor.

\subsection{Entropy production}
The entropy density is provided by
\begin{align}
T(x)\S(x) = p(x) - \mu_\sigma(x)\M_\sigma(x) + \E(x),
\end{align}
where $T(x)$ is the temperature and $\mu_\sigma(x)$ is the mass chemical potential for each spin component.
When $\S(x)$ is regarded as a local function of $\M_\sigma(x)$, $\E(x)$, and $a(x)$, it actually depends only on $a^{-d}(x)$ multiplied by a dimensionless function of $a^d(x)\M_\sigma(x)/m$ and $ma^{d+2}(x)\E(x)$.
Consequently, the partial derivative of $\S(x)$ with respect to $a(x)$ leads to
\begin{align}
& a(x)\frac{\d\S(x)}{\d a(x)} \notag\\
&= -d{\,\cdot\,}\S(x) + d{\,\cdot\,}\M_\sigma(x)\frac{\d\S(x)}{\d\M_\sigma(x)}
+ (d+2)\E(x)\frac{\d\S(x)}{\d\E(x)} \notag\\
&= -\frac{d{\,\cdot\,}p(x) - 2\E(x)}{T(x)},
\end{align}
so that together with Eq.~(\ref{eq:normal_pressure}) we obtain
\begin{align}
T(x)\frac{\d\S(x)}{\d a(x)} = -\frac{\C_\eq(x)}{m\Omega_{d-1}a^{d-1}(x)}.
\end{align}
This is the local extension of the thermodynamic identity known as the adiabatic relation~\cite{Tan:2008a,Tan:2008b,Tan:2008c,Braaten:2012} and the total differential of $\S(x)$ is now provided by
\begin{align}
T(x)d\S(x) &= -\mu_\sigma(x)d\M_\sigma(x) + d\E(x) \notag\\
&\quad - \frac{\C_\eq(x)}{m\Omega_{d-1}a^{d-1}(x)}da(x).
\end{align}

It is then straightforward to show that the above thermodynamic identities combined with the continuity equations (\ref{eq:continuity_mass})--(\ref{eq:continuity_energy}) and the constitutive relations in Sec.~\ref{sec:normal_constitutive} lead to the entropy production equation
\begin{align}
\d_t\S(x) + \d_i\biggl[\S(x)v_i(x) + \frac{q'_i(x)}{T(x)}\biggr] = \frac{\Phi(x)}{T(x)},
\end{align}
with the dissipation function provided by
\begin{align}
\Phi(x) &= -q'_i(x)\frac{\d_iT(x)}{T(x)} - \pi_{ij}(x)\d_iv_j(x) \notag\\
&\quad + \pi_{ii}(x)[\d_t\ln a(x) + v_k(x)\d_k\ln a(x)] + O(\d^3).
\end{align}
Here $q'_i(x)\equiv q_i(x)-\pi_{ij}(x)v_j(x)$ is the heat flux to be and we introduce the traceless part of the viscous stress tensor by $\pi'_{ij}(x)\equiv\pi_{ij}(x)-\delta_{ij}\pi_{kk}(x)/d$.
In order for the entropy production rate to be non-negative, the dissipative corrections up to first order in derivatives must be in the forms of
\begin{align}
q'_i(x) &= -\kappa(x)\d_iT(x) + O(\d^2), \\
\pi'_{ij}(x) &= -\eta(x)V_{ij}(x) + O(\d^2), \\
\pi_{ii}(x) &= -d{\,\cdot\,}\zeta(x)V_a(x) + O(\d^2),
\end{align}
where
\begin{align}\label{eq:strain_shear}
V_{ij}(x) \equiv \d_iv_j(x) + \d_jv_i(x) - \delta_{ij}\frac2d\,\d_kv_k(x)
\end{align}
is the usual shear strain rate but
\begin{align}\label{eq:strain_bulk}
V_a(x) &\equiv \d_kv_k(x) - d{\,\cdot\,}[\d_t\ln a(x) + v_k(x)\d_k\ln a(x)]
\end{align}
is the bulk strain rate modified by the spacetime-dependent scattering length.
Therefore, the dissipation function is found to be
\begin{align}\label{eq:normal_dissipation}
\Phi(x) &= \kappa(x)\frac{[\d_iT(x)]^2}{T(x)} + \frac{\eta(x)}{2}[V_{ij}(x)]^2 \notag\\
&\quad + \zeta(x)[V_a(x)]^2 + O(\d^3),
\end{align}
where the second law of thermodynamics is satisfied by imposing non-negativity on the thermal conductivity $\kappa(x)$, the shear viscosity $\eta(x)$, and the bulk viscosity $\zeta(x)$.
These transport coefficients depend on space and time because they are locally specified by $\M_\sigma(x)$, $\E(x)$, and $a(x)$.

We thus find that the spacetime-dependent scattering length enters the dissipation function partially as
\begin{align}
\Phi(x) \sim \frac{\zeta(x)}{a^2(x)}[\d_ta(x)]^2 \sim \zeta(x)a^2(x)\biggl[\d_t\frac1{a(x)}\biggr]^2.
\end{align}
In order for such a term to be nondivergent, the bulk viscosity must vanish at the slowest as
\begin{align}
\zeta(x) \sim a^2(x) \quad\text{for}\quad a(x) \to 0
\end{align}
and
\begin{align}\label{eq:normal_vanishing}
\zeta(x) \sim \frac1{a^2(x)} \quad\text{for}\quad a(x) \to \infty,
\end{align}
assuming that the hydrodynamics is applicable there.
In particular, the latter behavior proves to be consistent with the vanishing bulk viscosity of the unitary Fermi gas in a normal phase~\cite{Son:2007,Dusling:2013,Elliott:2014}.

\section{Hydrodynamics for superfluids}\label{sec:superfluids}
\subsection{Superfluid velocity}
The hydrodynamic equations for superfluids can also be constructed in a parallel way.
While the continuity equations remain the same because they follow from the operator identities in Eqs.~(\ref{eq:continuity_mass})--(\ref{eq:continuity_energy}), the constitutive relations must be modified by the presence of the superfluid velocity $u_i(x)\equiv[\d_i\theta(x)-A_i(x)]/m$.
Here $\theta(x)$ is the condensate phase normalized so as to transform as $\theta(x)\to\theta(x)+\chi(x)$ under the local gauge transformation in Eq.~(\ref{eq:gauge}) so that $u_i(x)$ is gauge invariant.
Its time evolution is governed by
\begin{align}\label{eq:super_velocity}
\d_tu_i(x) + \d_i\biggl[\frac{[u_j(x)]^2}{2} + \nu(x)\biggr] = \frac{F_{it}(x)}{m},
\end{align}
which follows from the fact that $m\nu(x)\equiv-[\d_t\theta(x)-A_t(x)]-m[u_j(x)]^2/2$ is a scalar field invariant under the Galilean transformation~\cite{Son:2006}.
The currently unknown potential $\nu(x)\equiv\bar\mu(x)+\mu'(x)$ is decomposed into the thermodynamic part $\bar\mu(x)$ and the dissipative correction $\mu'(x)\sim O(\d)$, both of which will be identified later in Sec.~\ref{sec:super_entropy}.

\subsection{Constitutive relations}\label{sec:super_constitutive}
In terms of the local thermodynamic variables, the normal fluid velocity $v_i(x)$, and the superfluid velocity $u_i(x)$, the constitutive relations for the conserved charge and current densities in Eqs.~(\ref{eq:operator_mass})--(\ref{eq:operator_energy-flux}) read
\begin{align}
\M_\sigma(x) = \M_\sigma^{(\n)}(x) + \M_\sigma^{(\s)}(x)
\end{align}
for the mass densities,
\begin{align}
\P_{\sigma i}(x) = \M_\sigma^{(\n)}(x)v_i(x) + \M_\sigma^{(\s)}(x)u_i(x)
\end{align}
for the momentum densities,
\begin{align}\label{eq:super_energy}
\H(x) = \E(x) + \P_i(x)u_i(x) - \frac{\M(x)}{2}[u_i(x)]^2
\end{align}
for the energy density,
\begin{align}\label{eq:super_stress}
\Pi_{ij}(x) &= p(x)\delta_{ij} + \M^{(\n)}(x)v_i(x)v_j(x) \notag\\
&\quad + \M^{(\s)}(x)u_i(x)u_j(x) + \pi_{ij}(x),
\end{align}
for the stress tensor, and
\begin{align}
& \Q_i(x) = [\H(x)+p(x)]v_i(x) \notag\\
& - \biggl[\mu_\sigma(x)\M_\sigma^{(\s)}(x) + \frac{\M^{(\s)}(x)}{2}[u_j(x)]^2\biggr]w_i(x) + q_i(x)
\end{align}
for the energy flux~\cite{Landau-Lifshitz}.
Here $\M^{(\n)}(x)\equiv\sum_\sigma\M_\sigma^{(\n)}(x)$ and $\M^{(\s)}(x)\equiv\sum_\sigma\M_\sigma^{(\s)}(x)$ are the normal fluid component and the superfluid component of the total mass density, respectively, and $w_i(x)\equiv v_i(x)-u_i(x)$ is their relative velocity.

In addition, by substituting the constitutive relations for the energy density and the stress tensor in Eqs.~(\ref{eq:super_energy}) and (\ref{eq:super_stress}) into the expectation value of the operator identity in Eq.~(\ref{eq:operator_stress-trace}), we obtain
\begin{align}\label{eq:super_stress-trace}
\frac{\C(x)}{m\Omega_{d-1}a^{d-2}(x)}
&= d{\,\cdot\,}p(x) - 2\E(x) + \M^{(\n)}(x)[w_i(x)]^2 \notag\\
&\quad + \pi_{ii}(x) + O(\d^2)
\end{align}
up to first order in derivatives.
Therefore, the contact density in local thermodynamic equilibrium is to be identified as
\begin{align}\label{eq:super_pressure}
\frac{\C_\eq(x)}{m\Omega_{d-1}a^{d-2}(x)}
\equiv d{\,\cdot\,}p(x) - 2\E(x) + \M^{(\n)}(x)[w_i(x)]^2,
\end{align}
which is the local pressure relation for the two-fluid hydrodynamics.
Here the equilibrium contact density $\C_\eq(x)$ is locally specified by $\M_\sigma(x)$, $\E(x)$, $\M^{(\n)}(x)w_i(x)$, and $a(x)$ via the thermodynamic equation of state and should be distinguished from the genuine contact density $\C(x)\equiv\Tr[\hat\C(x)\hat\rho]$ not necessarily in local thermodynamic equilibrium.
Its constitutive relation is thus found to be
\begin{align}\label{eq:super_contact}
\C(x) = \C_\eq(x) + m\Omega_{d-1}a^{d-2}(x)\pi_{ii}(x) + O(\d^2),
\end{align}
where the dissipative correction coincides with that to the stress tensor.

\subsection{Entropy production}\label{sec:super_entropy}
The entropy density is provided by~\cite{Landau-Lifshitz}
\begin{align}
T(x)\S(x) &= p(x) - \mu_\sigma(x)\M_\sigma(x) + \E(x) \notag\\
&\quad - \M^{(\n)}(x)[w_i(x)]^2.
\end{align}
When $\S(x)$ is regarded as a local function of $\M_\sigma(x)$, $\E(x)$, $\M^{(\n)}(x)w_i(x)$, and $a(x)$, it actually depends only on $a^{-d}(x)$ multiplied by a dimensionless function of $a^d(x)\M_\sigma(x)/m$, $ma^{d+2}(x)\E(x)$, and $a^{d+1}(x)\M^{(\n)}(x)w_i(x)$.
Consequently, the partial derivative of $\S(x)$ with respect to $a(x)$ leads to
\begin{align}
& a(x)\frac{\d\S(x)}{\d a(x)} \notag\\
&= -d{\,\cdot\,}\S(x) + d{\,\cdot\,}\M_\sigma(x)\frac{\d\S(x)}{\d\M_\sigma(x)}
+ (d+2)\E(x)\frac{\d\S(x)}{\d\E(x)} \notag\\
&\quad + (d+1)\M^{(\n)}(x)w_i(x)\frac{\d\S(x)}{\d[\M^{(\n)}(x)w_i(x)]} \notag\\
&= -\frac{d{\,\cdot\,}p(x) - 2\E(x) + \M^{(\n)}(x)[w_i(x)]^2}{T(x)},
\end{align}
so that together with Eq.~(\ref{eq:super_pressure}) we obtain
\begin{align}
T(x)\frac{\d\S(x)}{\d a(x)} = -\frac{\C_\eq(x)}{m\Omega_{d-1}a^{d-1}(x)}.
\end{align}
This is the local adiabatic relation for the two-fluid hydrodynamics and the total differential of $\S(x)$ is now provided by
\begin{align}
& T(x)d\S(x) = -\mu_\sigma(x)d\M_\sigma(x) + d\E(x) \notag\\
& - w_i(x)d[\M^{(\n)}(x)w_i(x)] - \frac{\C_\eq(x)}{m\Omega_{d-1}a^{d-1}(x)}da(x).
\end{align}

It is then straightforward to show that the above thermodynamic identities combined with the continuity equations (\ref{eq:continuity_mass})--(\ref{eq:continuity_energy}), Eq.~(\ref{eq:super_velocity}) and $\d_iu_j(x)-\d_ju_i(x)=-F_{ij}(x)/m$ for the superfluid velocity, and the constitutive relations in Sec.~\ref{sec:super_constitutive} lead to the entropy production equation
\begin{align}
\d_t\S(x) + \d_i\biggl[\S(x)v_i(x) + \frac{q'_i(x)}{T(x)}\biggr] = \frac{\Phi(x)}{T(x)},
\end{align}
with the dissipation function provided by
\begin{align}
\Phi(x) &= [\M_\sigma^{(\s)}(x)\d_i\mu_\sigma(x) - \M^{(\s)}(x)\d_i\bar\mu(x)]w_i(x) \notag\\
&\quad - q'_i(x)\frac{\d_iT(x)}{T(x)} - \pi_{ij}(x)\d_iv_j(x) \notag\\
&\quad + \pi_{ii}(x)[\d_t\ln a(x) + v_k(x)\d_k\ln a(x)] \notag\\
&\quad + \mu'(x)\d_i[\M^{(\s)}(x)w_i(x)] + O(\d^3).
\end{align}
Here $q'_i(x)\equiv q_i(x)-\pi_{ij}(x)v_j(x)+\mu'(x)\M^{(\s)}(x)w_i(x)$ is the heat flux to be and we introduce the traceless part of the viscous stress tensor by $\pi'_{ij}(x)\equiv\pi_{ij}(x)-\delta_{ij}\pi_{kk}(x)/d$.
In order for the entropy production rate to be non-negative, $\M_\sigma^{(\s)}(x)/\M^{(\s)}(x)$ must be constant over space so that
\begin{align}
\bar\mu(x) = \frac{\M_\sigma^{(\s)}(x)}{\M^{(\s)}(x)}\mu_\sigma(x)
\end{align}
is the mass chemical potential average weighted by the proportion of each spin component in the superfluid mass density.
In addition, the dissipative corrections up to first order in derivatives must be in the forms of
\begin{align}
q'_i(x) &= -\kappa(x)\d_iT(x) + O(\d^2), \\
\pi'_{ij}(x) &= -\eta(x)V_{ij}(x) + O(\d^2), \\
\pi_{ii}(x) &= -d{\,\cdot\,}\zeta_1(x)\d_i[\M^{(\s)}(x)w_i(x)] \notag\\
&\quad - d{\,\cdot\,}\zeta_2(x)V_a(x) + O(\d^2), \\
\mu'(x) &= \zeta_3(x)\d_i[\M^{(\s)}(x)w_i(x)] + \zeta_4(x)V_a(x) + O(\d^2),
\end{align}
where $V_{ij}(x)$ and $V_a(x)$ are the shear and bulk strain rates defined in Eqs.~(\ref{eq:strain_shear}) and (\ref{eq:strain_bulk}), respectively, and $\zeta_1(x)=\zeta_4(x)$ follows from the Onsager reciprocal relations.
Therefore, the dissipation function is found to be
\begin{align}\label{eq:super_dissipation}
\Phi(x) &= \kappa(x)\frac{[\d_iT(x)]^2}{T(x)} + \frac{\eta(x)}{2}[V_{ij}(x)]^2 \notag\\
&\quad + 2\zeta_1(x)V_a(x)\d_i[\M^{(\s)}(x)w_i(x)] + \zeta_2(x)[V_a(x)]^2 \notag\\
&\quad + \zeta_3(x)[\d_i[\M^{(\s)}(x)w_i(x)]]^2 + O(\d^3),
\end{align}
where the second law of thermodynamics is satisfied by imposing $\kappa(x)$, $\eta(x)$, $\zeta_2(x)$, $\zeta_3(x)\geq0$, and $\zeta_2(x)\zeta_3(x)\geq[\zeta_1(x)]^2$.
These transport coefficients depend on space and time because they are locally specified by $\M_\sigma(x)$, $\E(x)$, $\M^{(\n)}(x)w_i(x)$, and $a(x)$.

We thus find that the spacetime-dependent scattering length enters the dissipation function partially as
\begin{align}
\Phi(x) &\sim \frac{\zeta_1(x)}{a(x)}[\d_ta(x)] + \frac{\zeta_2(x)}{a^2(x)}[\d_ta(x)]^2 \notag\\
&\sim \zeta_1(x)a(x)\biggl[\d_t\frac1{a(x)}\biggr] + \zeta_2(x)a^2(x)\biggl[\d_t\frac1{a(x)}\biggr]^2.
\end{align}
In order for such terms to be nondivergent, the bulk viscosities must vanish at the slowest as
\begin{align}
\zeta_1(x) \sim a(x), \quad \zeta_2(x) \sim a^2(x) \quad\text{for}\quad a(x) \to 0
\end{align}
and
\begin{align}\label{eq:super_vanishing}
\zeta_1(x) \sim \frac1{a(x)}, \quad \zeta_2(x) \sim \frac1{a^2(x)} \quad\text{for}\quad a(x) \to \infty,
\end{align}
assuming that the hydrodynamics is applicable there.
In particular, the latter behaviors prove to be consistent with the vanishing bulk viscosities of the unitary Fermi gas in a superfluid phase~\cite{Son:2007,Escobedo:2009,Hou:2013}.

\section{Conclusion}\label{sec:conclusion}
The hydrodynamic equations consist of the continuity equations and the constitutive relations, which together with the equation of state and the transport coefficients provide a closed set of equations to govern long-time and long-distance physics of the correlated system out of thermodynamic equilibrium.
In this paper, we constructed the hydrodynamic equations with the spacetime-dependent scattering length and showed that it enters not only the momentum and energy continuity equations as the external sources [Eqs.~(\ref{eq:continuity_momentum}) and (\ref{eq:continuity_energy})], but also the constitutive relations via the modified bulk strain rate [Eq.~(\ref{eq:strain_bulk})] in both normal and superfluid phases.
While the modified bulk strain rate is uniquely identified by imposing the second law of thermodynamics, the resulting formula is intuitively understandable, i.e., the expansion (contraction) of fluid volume at a rate $\d_kv_k(x)$ is equivalent to the contraction (expansion) of scattering length at a rate $\d_kv_k(x)/d$ because no other reference scales exist in contact interactions.
In addition, $\d_t\ln a(x)$ must be the material derivative accompanied by $v_k(x)\d_k\ln a(x)$ to ensure the Galilean invariance.
As a consequence, the spacetime-dependent scattering length is naturally coupled with the bulk viscosities.

It is also worthwhile to remark that our formula in Eq.~(\ref{eq:strain_bulk}) is consistent with the conformal invariance in curved space~\cite{Son:2006,Son:2007,Chao:2012}.
Even though the conformal invariance is explicitly broken by the presence of a nonzero and finite scattering length, it is formally recovered by regarding the scattering length as a spacetime-dependent spurion field with conformal dimension $\Delta_a=-1/2$.
The bulk strain rate that transforms as a scalar field under the nonrelativistic diffeomorphism,
\begin{align}
\nabla_kv^k(x) + \d_t\ln\!\sqrt{g(x)},
\end{align}
was found to be incompatible with the conformal invariance because its conformal transformation involves an undesired term of $(d/2)\ddot\beta(t)$~\cite{Son:2007}.%
\footnote{Here $\nabla_i$ is the covariant derivative with respect to an external metric $g_{ij}(x)$ and its determinant is denoted by $g(x)\equiv\det[g_{ij}(x)]$.}
However, such a term can be eliminated by modifying the bulk strain rate as
\begin{align}
\nabla_kv^k(x) + \d_t\ln\!\sqrt{g(x)} - d{\,\cdot\,}[\d_t\ln a(x) + v^k(x)\d_k\ln a(x)],
\end{align}
which is the unique combination allowed by the diffeomorphism and conformal invariance in the viscous stress tensor with nonvanishing bulk viscosity coefficients.
In flat space, Eq.~(\ref{eq:strain_bulk}) is readily obtained.

Finally, physical implications are to be extracted from our findings.
As we already showed in Eqs.~(\ref{eq:normal_vanishing}) and (\ref{eq:super_vanishing}), the vanishing bulk viscosities can be reproduced for the unitary Fermi gas in both normal and superfluid phases~\cite{Son:2007,Escobedo:2009,Hou:2013,Dusling:2013,Elliott:2014}.
In addition, let us consider for simplicity a uniform system where the fluid is at rest but the scattering length is slowly varied over time.
According to Eqs.~(\ref{eq:normal_contact}) and (\ref{eq:super_contact}) (see also Appendix~\ref{sec:appendix}), the dissipative correction to the contact density proves to be proportional to the bulk viscosity,
\begin{align}\label{eq:production_contact}
\C(t) = \C_\eq(t) + m\Omega_{d-1}d^2{\,\cdot\,}\zeta(t)a^{d-3}(t)\dot{a}(t) + O(\dot{a}^2),
\end{align}
which combined with Eq.~(\ref{eq:continuity_energy}) leads to the energy density produced at the rate of
\begin{align}\label{eq:production_energy}
\dot\H(t) = \frac{\C_\eq(t)}{m\Omega_{d-1}a^{d-1}(t)}\dot{a}(t)
+ \frac{d^2{\,\cdot\,}\zeta(t)}{a^2(t)}\dot{a}^2(t) + O(\dot{a}^3).
\end{align}
Similarly, according to Eqs.~(\ref{eq:normal_dissipation}) and (\ref{eq:super_dissipation}), the entropy density proves to be produced even without fluid flows at the rate of
\begin{align}\label{eq:production_entropy}
T(t)\dot\S(t) = \frac{d^2{\,\cdot\,}\zeta(t)}{a^2(t)}\dot{a}^2(t) + O(\dot{a}^3),
\end{align}
where $\zeta(x)$ in the normal phase is replaced with $\zeta_2(x)$ in the superfluid phase.
Therefore, we find that the leading (subleading) contribution to the entropy (energy) density production due to the time-dependent scattering length is proportional to the bulk viscosity, which may be useful as a novel probe to measure the bulk viscosity in ultracold-atom experiments.

\acknowledgments
The authors thank Yoshimasa Hidaka, Masaru Hongo, Munekazu Horikoshi, and Yuta Sekino for valuable discussions.
This work was supported by JSPS KAKENHI Grants No.~JP15K17727 and No.~JP15H05855.
One of the authors (K.F.) was also supported by International Research Center for Nanoscience and Quantum Physics, Tokyo Institute of Technology and by RIKEN Junior Research Associate Program.

\appendix
\section{Microscopic derivation of Eqs.~(\ref{eq:production_contact})--(\ref{eq:production_entropy})}\label{sec:appendix}
While Eqs.~(\ref{eq:production_contact})--(\ref{eq:production_entropy}) were derived on the ground of hydrodynamics, they can also be confirmed microscopically by employing linear response theory.
Let us consider a uniform system in the rest frame, $A_\mu(x)\to0$, which is perturbed by a time-dependent scattering length varied slightly from its constant value as
\begin{align}
a(t) = a_0 + \delta a(t).
\end{align}
Consequently, the Hamiltonian and contact operators in the Schr\"odinger picture vary as
\begin{align}
\hat{H}(t) &\equiv \int\!d^d\x\,\hat\H(x)
= \hat{H}_0 + \frac{\hat{C}_0}{m\Omega_{d-1}a_0^{d-1}}\delta a(t), \\
\hat{C}(t) &\equiv \int\!d^d\x\,\hat\C(x)
= \hat{C}_0 + \frac{\d\hat{C}_0}{\d a_0}\delta a(t)
\end{align}
and the density matrix as
\begin{align}
\hat\rho(t) = \hat\rho_0 + \delta\hat\rho(t),
\end{align}
where $\hat\rho_0\equiv e^{-\hat{H}_0/T}/\Tr[e^{-\hat{H}_0/T}]$ is the equilibrium density matrix and
\begin{align}
\delta\hat\rho(t) &= -\frac{i}{m\Omega_{d-1}a_0^{d-1}}\int\!dt'
e^{-i\hat{H}_0(t-t')}[\hat{C}_0,\hat\rho_0]e^{i\hat{H}_0(t-t')} \notag\\
&\qquad \times \theta(t-t')\delta a(t')
\end{align}
up to first order in the perturbation.
The expectation value of the contact operator is thus provided by
\begin{align}\label{eq:appendix_contact-def}
& C(t) \equiv \Tr[\hat{C}(t)\hat\rho(t)] \notag\\
&= \Tr[\hat{C}_0\hat\rho_0] + \Tr\biggl[\frac{\d\hat{C}_0}{\d a_0}\hat\rho_0\biggr]\delta a(t)
+ \Tr[\hat{C}_0\delta\hat\rho(t)] + O(\delta^2a),
\end{align}
where the third term is expressed as
\begin{align}
\Tr[\hat{C}_0\delta\hat\rho(t)] &= -\frac{i}{m\Omega_{d-1}a_0^{d-1}}\int\!dt'
\Tr[[\hat{C}_0(t),\hat{C}_0(t')]\hat\rho_0] \notag\\
&\qquad \times \theta(t-t')\delta a(t')
\end{align}
in terms of the unperturbed contact operator $\hat{C}_0(t)\equiv e^{i\hat{H}_0t}\hat{C}_0e^{-i\hat{H}_0t}$ in the interaction picture.

On the other hand, by setting $a(x)\to a_0$ in Eq.~(\ref{eq:operator_stress-trace}) and integrating it over space, the operator identity in the interaction picture follows as
\begin{align}
\int\!d^d\x\,\hat\Pi_{ii}^0(t,\x) = 2\hat{H}_0 + \frac{\hat{C}_0(t)}{m\Omega_{d-1}a_0^{d-2}},
\end{align}
so that we obtain
\begin{align}
& \Tr[\hat{C}_0\delta\hat\rho(t)]
= -im\Omega_{d-1}a_0^{d-3}\int\!d^d\x\int\!d^d\x'\!\int\!dt' \notag\\
&\qquad \times \Tr[[\hat\Pi_{ii}^0(t,\x),\hat\Pi_{jj}^0(t',\x')]\hat\rho_0]\theta(t-t')\delta a(t').
\end{align}
Its Fourier transformation, the Kubo formula for the bulk viscosity in terms of the stress-stress response function~\cite{Bradlyn:2012},
\begin{align}
\zeta_0 \equiv \lim_{\omega\to0}\int\!d^d\x\int\!dt\,\frac{e^{i\omega t}-1}{d^2{\,\cdot\,}\omega}
\Tr[[\hat\Pi_{ii}^0(t,\x),\hat\Pi_{jj}^0(0,\0)]\hat\rho_0]\theta(t),
\end{align}
and then the inverse Fourier transformation lead to
\begin{align}
& \Tr[\hat{C}_0\delta\hat\rho(t)]
= -im\Omega_{d-1}a_0^{d-3}\int\!d^d\x\int\!d^d\x'\!\int\!dt' \notag\\
&\qquad \times \Tr[[\hat\Pi_{ii}^0(t,\x),\hat\Pi_{jj}^0(t',\x')]\hat\rho_0]\theta(t-t')\delta a(t) \notag\\
&\quad + Vm\Omega_{d-1}a_0^{d-3}d^2{\,\cdot\,}\zeta_0\,\delta\dot{a}(t) + O(\delta\ddot{a}) \notag\\
&= \Tr\biggl[\hat{C}_0\frac{\d\hat\rho_0}{\d a_0}\biggr]\delta a(t)
+ Vm\Omega_{d-1}d^2{\,\cdot\,}\zeta_0\,a_0^{d-3}\delta\dot{a}(t) + O(\delta\ddot{a}).
\end{align}
Therefore, the contact in Eq.~(\ref{eq:appendix_contact-def}) proves to be
\begin{align}\label{eq:appendix_contact}
C(t) &= \Tr[\hat{C}_0\hat\rho_0] + \Tr\biggl[\frac{\d\hat{C}_0}{\d a_0}\hat\rho_0\biggr]\delta a(t)
+ \Tr\biggl[\hat{C}_0\frac{\d\hat\rho_0}{\d a_0}\biggr]\delta a(t) \notag\\
&\quad + Vm\Omega_{d-1}d^2{\,\cdot\,}\zeta_0\,a_0^{d-3}\delta\dot{a}(t) + O(\delta^2a,\delta\ddot{a}) \notag\\
&= C_\eq[a(t)] + Vm\Omega_{d-1}d^2{\,\cdot\,}\zeta_0\,a^{d-3}(t)\dot{a}(t) + O(\delta^2a,\delta\ddot{a}),
\end{align}
where $V$ is the volume and $C_\eq[a_0]\equiv\Tr[\hat{C}_0\hat\rho_0]$ is the contact for the constant scattering length $a_0$ in thermodynamic equilibrium.

Finally, the expectation value of the energy continuity equation (\ref{eq:continuity_energy}) integrated over space leads to the energy production at the rate of
\begin{align}\label{eq:appendix_energy}
\dot{H}(t) &= \frac{C(t)}{m\Omega_{d-1}a^{d-1}(t)}\dot{a}(t) \notag\\
&= \frac{C_\eq[a(t)]}{m\Omega_{d-1}a^{d-1}(t)}\dot{a}(t)
+ V\frac{d^2{\,\cdot\,}\zeta_0}{a^2(t)}\dot{a}^2(t) + O(\delta^3\dot{a},\delta^2\dddot{a}),
\end{align}
where the former equality is also known as the dynamic sweep theorem~\cite{Tan:2008a,Tan:2008b,Tan:2008c,Braaten:2012}.
The entropy production at the rate of
\begin{align}\label{eq:appendix_entropy}
T\dot{S}(t) &= \dot{H}(t) - \frac{C_\eq[a(t)]}{m\Omega_{d-1}a^{d-1}(t)}\dot{a}(t) \notag\\
&= V\frac{d^2{\,\cdot\,}\zeta_0}{a^2(t)}\dot{a}^2(t) + O(\delta^3\dot{a},\delta^2\dddot{a})
\end{align}
then follows from the thermodynamic identity combined with the adiabatic relation~\cite{Tan:2008a,Tan:2008b,Tan:2008c,Braaten:2012}.
We thus find that Eqs.~(\ref{eq:appendix_contact}), (\ref{eq:appendix_energy}), and (\ref{eq:appendix_entropy}) divided by the volume reproduce Eqs.~(\ref{eq:production_contact}), (\ref{eq:production_energy}), and (\ref{eq:production_entropy}), respectively, from a microscopic perspective without relying on the hydrodynamics.

\end{document}